\documentclass[aps,prl,twocolumn,preprintnumbers]{revtex4}
\usepackage{amssymb}
\usepackage{times}
\usepackage{epsfig}
\usepackage{amsmath}
\usepackage{bm}
\usepackage{pstricks,fancybox}

\begin{document}

\title{Solvable Dynamics of Coupled High-Dimensional Generalized Limit-Cycle Oscillators}

\author{Wei Zou$^{1}$}
\email{weizou83@gmail.com}
\author{Sujuan He$^{1}$}
\author{D.V. Senthilkumar$^{2}$}
\author{J\"{u}rgen Kurths$^{3,4}$}

\affiliation{
\\$^{1}$ School of Mathematical Sciences, South China Normal University, Guangzhou 510631, China
\\$^{2}$ School of Physics, Indian Institute of Science Education and Research, Thiruvananthapuram 695551, Kerala, India
\\ $^{3}$ Potsdam Institute for Climate Impact Research, Telegraphenberg, Potsdam D-14415, Germany
\\$^{4}$Institute of Physics, Humboldt University Berlin, Berlin D-12489, Germany}
\date{\today}

\begin{abstract}
We introduce a new model consisting of globally coupled high-dimensional generalized limit-cycle oscillators,
which explicitly incorporates the role of amplitude dynamics of individual units in the collective dynamics.
In the limit of weak coupling, our model reduces to the $D$-dimensional Kuramoto phase model,
akin to a similar classic construction of the well-known Kuramoto phase model from weakly
coupled two-dimensional limit-cycle oscillators. 
For the practically important case of $D=3$, the incoherence of the model is rigorously proved to be stable for negative coupling $(K<0)$ but unstable for positive coupling $(K>0)$;
the locked states are shown to exist if $K>0$; in particular, the onset of amplitude death is theoretically predicted.
For $D\geq2$, the discrete and continuous spectra for both locked states and amplitude death are governed by two general formulas.
Our proposed $D$-dimensional model is physically more reasonable, because it is no longer constrained by fixed amplitude dynamics,
which puts the recent studies of the $D$-dimensional Kuramoto phase model
on a stronger footing by providing a more general framework for $D$-dimensional limit-cycle oscillators.
\end{abstract}

\maketitle
\newpage
\bigskip

Self-organization of collective behavior from interacting units is ubiquitous in nature \cite{Book1,Book2,Book3,Book4}, which
can be qualitatively and quantitatively explored by employing models of coupled nonlinear oscillators \cite{Book5,Vicsek1995,Chate2004}.
Among them, the celebrated Kuramoto model \cite{Kuramoto_model} has served as a paradigm for the study of
synchronization in wide disciplines ranging from physics, biology to engineering \cite{Strogatz2000,Acebron2005,Kiss2002}.
For better understanding the mechanisms of synchronization,
Ritort \cite{Ritort1998} introduced a solvable model of interacting random tops incorporating the
orientational degree of freedom, which in fact extends the Kuramoto model with noise to three dimensions \cite{Zheng2021}.
Later on, the $D$-dimensional Kuramoto model has been further proposed \cite{Saber2006,Zhu2013,Tanaka2014}, where all the individual oscillators (agents) are
interpreted as $D$-dimensional unit vectors, rotating on the surface of the $D$-dimensional sphere.
Quite recently, Chandra, Girvan, and Ott \cite{Chandra2019} systematically examined the dynamics of the $D$-dimensional generalized Kuramoto model with heterogeneous natural rotations;
in particular, they unveiled that the nature of phase transition for the generalized Kuramoto model with the odd number of dimensions is remarkably different from that in even dimensions.
Since then, there has been a burst of appealing works devoted to the study of the $D$-dimensional generalized Kuramoto model and its variants \cite{Chandra2019b,Chandra2019c,Dai2020,Kovalenko2021,Dai2021,Lipton2021,Barioni2021,Barioni2021b}.

However, amplitude dynamics of individual units has not been taken into account in the above studies. This strongly limits the applicability of the model,
as the amplitude degree of freedom generally plays a key role in determining collective dynamics
of strongly coupled systems, examples including a flock of birds, a school of fish, a swarm of flying drones or insects \cite{Katz2011,Vicsek2012,Keeffe2017,Sumpter2010}, etc.
Phase-amplitude models are deemed to capture common neuroimaging metrics more
accurately and are important to quantify anaesthetised brain states \cite{Fagerholm2020}.
A full representation of the phase and amplitude coordinates is of quite relevance for
understanding bifurcations of high-dimensional nonlinear systems beyond the weak coupling limit \cite{Wilson2019}.

To resolve this limitation, in this Letter, we propose a new model of globally coupled $D$-dimensional generalized
limit-cycle oscillators, which explicitly incorporates both phase and amplitude dynamics of individual units.
Our model includes the $D$-dimensional Kuramoto phase model as a special case in the weak coupling limit.
Of particular interest, we show that our model for the practically important case of $D=3$ is solvable in the thermodynamic limit,
which provides a new paradigmatic example of analytically tractable models.
The high-dimensional model proposed in this Letter is expected to better capture emergent dynamics in diverse physical and biological systems comprised of interacting units with natural magnetic
moments \cite{Heinrich2003,Klingler2018,Young2020}, such as strongly coupled magnetic particles \cite{Yan2012,Snezhko2011,Martin2013} and microfluidic mixtures of active spinners \cite{Nguyen2014,Zuiden2016}.

The model consists of a system of $N$ globally coupled $D$-dimensional vectors $\vec{r}_i\in \mathbb{R}^{D}$ described by
\begin{eqnarray}
\label{DSL}
\frac{d\vec{r}_i}{dt}=(1-|\vec{r}_i|^2)\vec{r}_i+\textbf{W}_i\vec{r}_i+\frac{K}{N}\sum_{\substack{j=1}}^N(\vec{r}_j-\vec{r}_i)
\end{eqnarray}
with $i=1,2,...,N$, where $K$ is the coupling strength, $\textbf{W}_i$ is a real $D\times D$ antisymmetric (skew-symmetric) matrix with $D(D-1)/2$
independent components, which can be physically interpreted as the natural rotation of the $i$th agent \cite{MatrixW}.
Each $\textbf{W}_i$ is coded by the vector $\vec{\omega}_i=(\omega_{1i},\omega_{2i},...,\omega_{\frac{D(D-1)}{2}i})^{T}$ drawn from a normalized distribution $G(\vec{\omega})$.
In the limit of $K \rightarrow 0$, (\ref{DSL}) degenerates to the $D$-dimensional Kuramoto model \cite{SM}.
\begin{figure}[!h]
\begin{raggedright}
\centering\includegraphics[width=0.9\columnwidth]{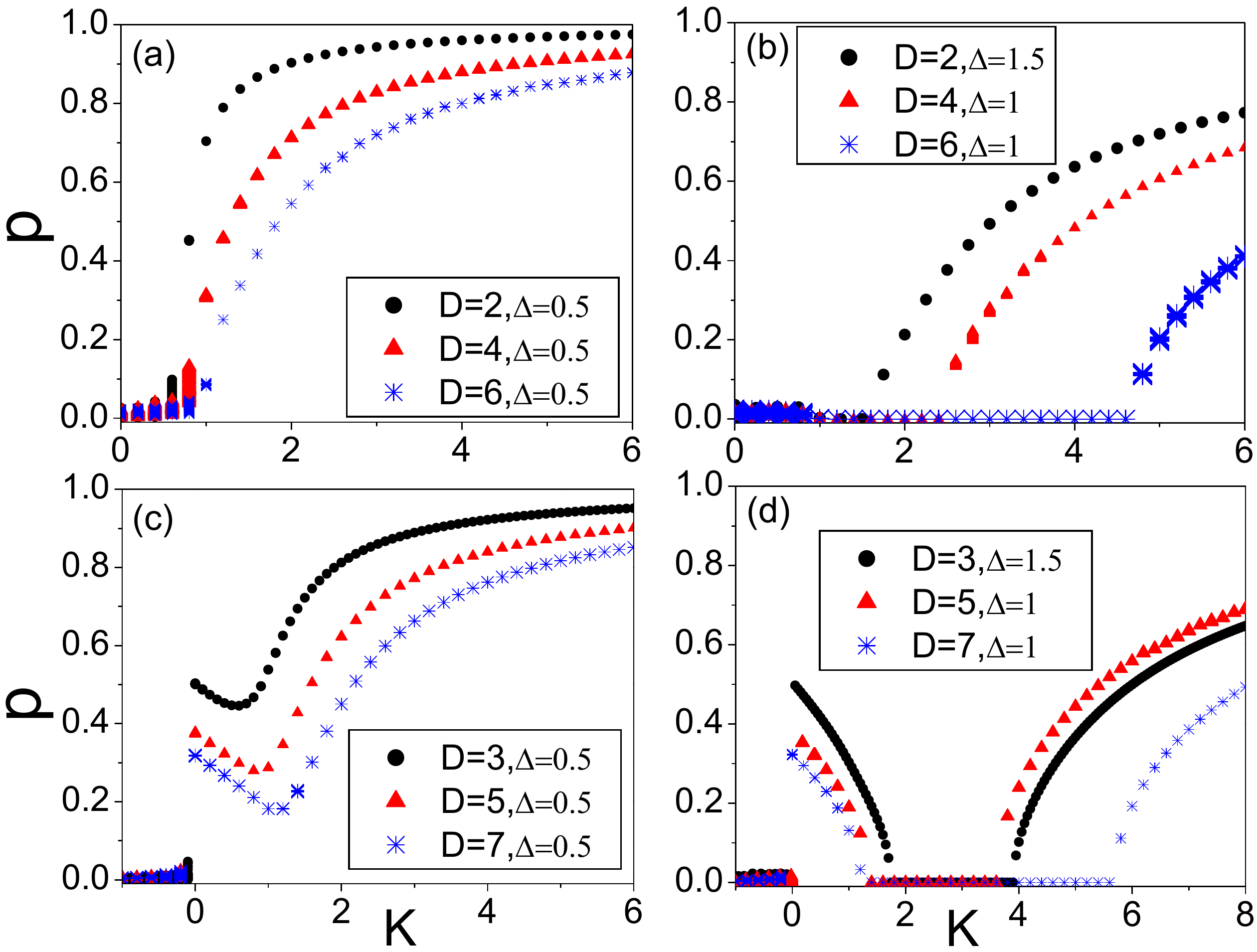}
\par\end{raggedright}
\centering \caption{ The magnitude $p$ of $\vec{p}$ vs $K$ for the system (\ref{DSL}) with $N=5000$
for $D=2-7$. Each element of $\vec{\omega}_i=(\omega_{1i},\omega_{2i},...,\omega_{\frac{D(D-1)}{2}i})^{T}$ is sampled randomly
according to ${Norm}(0,\Delta^2)$. A small value of $\Delta=0.5$ is used in the left column, while comparatively large values of $\Delta$ are employed in the right column.}\label{fig1}
\end{figure}
For $D=2$, (\ref{DSL}) reduces to the model addressed in \cite{Shiino1989,Matthews1990,Mirollo1990,Matthews1991,Ermentrout1990},
where the single uncoupled unit (coined as the Stuart-Landau oscillator) represents a canonical form near a supercritical Hopf bifurcation \cite{{Book1}}.
For $K\rightarrow 0$, the first-order phase reduction to (\ref{DSL}) with $D=2$ results in the classic Kuramoto model \cite{Kuramoto_model};
the second-order phase-reduction approach leads to the {\it enlarged} Kuramoto model \cite{Leon2019,Leon2022}.

Collective behavior in the model (\ref{DSL}) can be conveniently described by an order parameter $\vec{p}=\frac{1}{N}\sum_{\substack{i=1}}^N\vec{r}_i$,
where $p=|\vec{p}|$ measures the degree of collective synchronization.
Depending on $K$ and the spread of $\vec{\omega}_i$, the model (\ref{DSL}) exhibits three types of steady behaviors: {\it incoherence}, {\it amplitude death}, and {\it locking},
in which the system evolves to statistical steady states, characterized by a stationary distribution
of oscillators in the phase space and a constant $\vec{p}$.

Figure \ref{fig1} shows numerical observations of $p$ vs $K$ for the system  (\ref{DSL}) with $D=2-7$ and
$N=5000$. The $D(D-1)/2$ upper-triangular elements of $\textbf{W}_i$ are chosen randomly according to a normal distribution with zero mean and the standard
deviation $\Delta$, i.e., $\omega_{ji} \sim{Norm}(0,\Delta^2)$ for $j=1,2,...,D(D-1)/2$, while the corresponding lower-triangular elements are set to make $\textbf{W}_i$ to be an antisymmetric matrix.
With the above choices of $\textbf{W}_i$'s, $\vec{p}$ always asymptotically reaches an equilibrium. 
We find that for all odd $D$, the transition from incoherence to coherence occurs discontinuously as $K$ increases through zero (i.e., $K_c=0$). In contrast in the even $D$ case
the phase transition takes place continuously at $K_c>0$. A similar difference in the nature of the phase transition between odd and even dimensions has been previously established in
the $D$-dimensional Kuramoto model by Chandra {\it et al.} \cite{Chandra2019}. However, our high-dimensional model (\ref{DSL}) is no longer constrained by fixed
amplitude dynamics, which may render the emergence of richer dynamics, such as amplitude death as shown in the right column of Fig. \ref{fig1} for large values of $\Delta$.
To predict the onset of emergent dynamics observed in Fig. \ref{fig1}, we now conduct a theoretical analysis of the model (\ref{DSL}) for $D\geq3$.


{\it i) Stability of incoherence.}  For $D=3$, the incoherent state refers to that each oscillator rotates rigidly around the vector $\hat{\omega}_i$ at its natural rotation rate $\omega_i$ on the sphere with radius
of $\sqrt{1-K}$ ($K<1$), and meanwhile $\vec{p}=\vec{0}$ (i.e., $p=0$) holds at all times. Strictly speaking, the incoherent solution exists only when $N\rightarrow\infty$,
for which $\vec{p}$ becomes
\begin{eqnarray}
\label{orderparab}
\vec{p}=\int_{\mathbb{R}^3}\int_{\mathbb{R}^3}  \vec{r}   F(\vec{r},\vec{\omega},t)  G(\vec{\omega}) d\vec{\omega} d\vec{r},
\end{eqnarray}
where $F(\vec{r},\vec{\omega},t)$ represents a time dependent joint density of $\vec{r}$ and $\vec{\omega}$, satisfying the continuity equation
\begin{eqnarray}
\label{ce}
\frac{\partial F}{\partial t}+\frac{\partial (F\dot{r})}{\partial r}+\frac{1}{\sin \theta}\frac{\partial (F\sin \theta \dot{\theta})}{\partial \theta}+\frac{\partial (F\dot{\phi})}{\partial \phi}=0,
\end{eqnarray}
where $\dot{r}$, $ \dot{\theta}$, and $ \dot{\phi}$ can be calculated directly from (\ref{DSL}) via
introducing the spherical coordinates $\vec{r}=r \hat{r}$ with $\hat{r}=(\sin \theta \cos\phi,\sin \theta \sin\phi,\cos\theta)^T$.
For the incoherence, the oscillators are uniformly distributed on the sphere with the radius $\sqrt{1-K}$ for
each $\vec{\omega}$, then the corresponding density is $F_0=\frac{\delta(r-a)}{4\pi}$ with $a^2=1-K$.

To analyze the linear stability of incoherence, we introduce a small perturbation to the incoherent solution as
$F=F_0+\varepsilon e^{st}\xi(r,\theta,\phi,\vec{\omega})$ $(0<\varepsilon\ll 1)$ \cite{perturbations}. Then the perturbed order parameter is
\begin{eqnarray}
\label{orderparac}
\vec{p}=\varepsilon e^{st}\int_{\mathbb{R}^3}\int_{\mathbb{R}^3}  \vec{r}   \xi  G(\vec{\omega}) d\vec{\omega} d\vec{r}\triangleq\varepsilon e^{st} \langle \vec{\xi} \rangle .
\end{eqnarray}
Inserting the perturbed density into (\ref{ce}), we derive that
\begin{eqnarray}
\label{dispersion}
s\xi+\frac{\partial [(1-K-r^2)r\xi]}{\partial r}+\omega_{\phi}\frac{\partial \xi}{\partial \theta}-\frac{\omega_{\theta}}{\sin \theta}\frac{\partial \xi}{\partial \phi} \notag \\ =\frac{K\delta(r-a)}{2\pi a}
\langle \vec{\xi}  \rangle \cdot  \hat{r}-\frac{K\delta^{'}(r-a)}{4\pi} \langle \vec{\xi} \rangle \cdot  \hat{r},
\end{eqnarray}
where $\omega_{\theta}=\vec{\omega} \cdot \hat{\theta}$ with $\hat{\theta}=(\cos \theta\cos\phi, \cos \theta \sin \phi, -\sin \theta)^T$, $\omega_{\phi}=\vec{\omega} \cdot \hat{\phi}$ with $\hat{\phi}=(-\sin \phi, \cos \phi, 0)^T$, and $\delta^{'}(r-a)=d \delta(r-a)/ d r$. The solution of $\xi$ for (\ref{dispersion}) has the following form
\begin{eqnarray}
\label{solution}
\xi=\frac{K\delta(r-a)}{2\pi a}\vec{A} \cdot  \hat{r}-\frac{K\delta^{'}(r-a)}{4\pi} \vec{B} \cdot  \hat{r}
\end{eqnarray}
with $\vec{A}=(s\textbf{I}-\textbf{W})^{-1}\langle \vec{\xi} \rangle$ and $\vec{B}=[(s+2a^2)\textbf{I}-\textbf{W}]^{-1}\langle \vec{\xi} \rangle$.
Substituting $\xi$ from (\ref{solution}) into (\ref{orderparac}),
the dispersion relation for $s$ is obtained as \cite{SM}.
\begin{eqnarray}
\label{dispersiona}
\text{det}\left  (\textbf{I}-\frac{2K}{3}\textbf{J}(s)-\frac{K}{3}\textbf{J}(s+2a^2) \right  )=0,
\end{eqnarray}
where $\textbf{J}(s)=\int_{\mathbb{R}^3}(s\textbf{I}-\textbf{W})^{-1}  G(\vec{\omega}) d\vec{\omega}$.
If (\ref{dispersiona}) has one root with a positive real part, the incoherent state is unstable.

Assuming that the rotation directions of individual oscillators are isotropically distributed on the unit sphere,
and independent of the distribution of the rotation magnitudes $g(\omega)$, one can write
$G(\vec\omega)=g(\omega)U(\hat{\omega})$, where $U(\hat{\omega})=\frac{1}{4\pi}$.
With the above form of $G(\vec\omega)$,  we calculate that \cite{SM}
\begin{eqnarray}
\label{dispersionc}
\textbf{J}(s)=\left(\frac{1}{3s}+\frac{2s}{3}\int_0^{+\infty} \frac{g(\omega)}{s^2+\omega^2}d\omega\right)\textbf{I}\triangleq h(s)\textbf{I}.
\end{eqnarray}
The dispersion relation in (\ref{dispersiona}) finally reduces to
\begin{eqnarray}
\label{dispersiond}
1-\frac{2K}{3}h(s)-\frac{K}{3}h(s+2a^2)=0.
\end{eqnarray}
For $K\to 0$, $s\to 0$, the behavior of $s$ in (\ref{dispersiond}) for $K$ around zero is represented by $s=\frac{2}{9}K$ \cite{SM}.
Thus, we can ascertain that the incoherent state will be stable for $K<0$ and unstable for $K>0$, which is valid independent of $g(\omega)$. For $D=3$, we have proved that the incoherence loses
its stability at $K_c=0$, which is exactly the same as that of the $3$D Kuramoto model \cite{Chandra2019}, in turn confirming that our
model (\ref{DSL}) reduces to the $D$-dimensional Kuramoto model in the limit $K\rightarrow0$.

{\it ii) Stability of locking and amplitude death.}  Locked states correspond to fixed points of (\ref{DSL}), for which $\vec{p}$ is a constant vector with $p>0$.
In contrast, amplitude death refers to the coupling-induced stabilization of ${\vec r}_i=\vec{0}$, for which $p=0$.
Theoretically, the stability of locked states and amplitude death can be analyzed at the same time.
For $N \to \infty$, in the locked state, the position $\vec{r}_i$ of the $i$th oscillator is determined by its natural rotation $\textbf{W}_i$. Therefore,
$\vec{r}$ is regarded as a function of $\textbf{W}$ instead of the subscript $i$, which obeys
\begin{eqnarray}
\label{SL3D}
\frac{d \vec{r}_F}{dt}=(1-K-|\vec{r}_F|^2)\vec{r}_F+\textbf{W}\vec{r}_F+K\vec{p}_F=0
\end{eqnarray}
with the subscript $F$ indicating that the oscillator is at a fixed point.  The order parameter is then written by
$\vec{p}_F=\int_{\mathbb{R}^{D(D-1)/2}}\vec{r}_FG(\vec{\omega}) d\vec{\omega}$, whose magnitude $p_F$ satisfies \cite{SM}
\begin{eqnarray}
\label{SL3Dfpsc}
Kp_F^2=\int_{\mathbb{R}^{D(D-1)/2}}(|\vec{r}_F|^2-1+K)|\vec{r}_F|^2G(\vec{\omega}) d\vec{\omega},
\end{eqnarray}
which holds for all $D\geq2$.
For $D=3$, $\textbf{W}\vec{r}_F=\vec{\omega}\times\vec{r}_F$, (\ref{SL3D}) can be solved to obtain \cite{SM}
\begin{eqnarray}
\label{SL3Dfps}
\vec{r}_F=\frac{K}{(\mu^2+\omega^2)}\left( \mu\vec{p}_F+\nu \vec{\omega} +\vec{\omega}\times \vec{p}_F\right)
\end{eqnarray}
with $\mu=|\vec{r}_F|^2-1+K$ and $\nu=(\vec{p}_F\cdot \vec{\omega})/\mu$, where $\mu$ and $\nu$ obey
$(\mu+1-K)(\mu^2+\omega^2)=K^2(p_F^2+\nu^2)$.
From (\ref{SL3Dfps}), $\vec{r}_F$ always exist for $D=3$ once if $K>0$,
in contrast to $D=2$ for which locked state exists only for sufficiently large $K$ \cite{Matthews1990,Mirollo1990,Matthews1991}.

To determine the stability of locked states in the infinite-$N$ limit, one has to consider both the discrete and
the continuous spectrum of the linearized system of (\ref{SL3D}) around $\vec{r}_F$. For $D\geq2$ and $N\rightarrow\infty$, we find the continuous and
the discrete spectrum given by \cite{SM}:
\begin{eqnarray}
\label{LScs}
\text{det}\left  (s\textbf{I}-\textbf{M} \right )=0,
\end{eqnarray}
and
\begin{eqnarray}
\label{LSds}
\text{det}\left  (\textbf{I}-K\int_{\mathbb{R}^{D(D-1)/2}}(s\textbf{I}-\textbf{M})^{-1}  G(\vec{\omega}) d\vec{\omega}\right )=0,
\end{eqnarray} where $\textbf{M}=\textbf{W}-2\vec{r}_F\vec{r}_F^{\ T}+(1-|\vec{r}_F|^2-K)\textbf{I}$.
The locked solutions are stable if both (\ref{LScs}) and (\ref{LSds}) have only roots $s$ with $\text{Re}(s)<0$.

By setting $\vec{r}_F=\vec{0}$, (\ref{LSds}) and (\ref{LScs}) yield the discrete and
the continuous spectrum governing the stability of amplitude death for $N\rightarrow\infty$,
which can also be obtained by performing a stability analysis of (\ref{DSL}) around ${\vec r}_i=\vec{0}$ \cite{SM}.
The continuous spectrum can be proved to be stable if $K>1$ for all $D\geq2$ \cite{ADcsstable}, whereas its discrete spectrum
cannot be worked out explicitly for a general $G(\vec{\omega})$. Here, we analytically solves the stability of amplitude death
for all $D\geq2$, in contrast to the earlier works \cite{Mirollo1990,Ermentrout1990} confined to $D=2$.

\begin{figure}
\begin{center}
\centering\includegraphics[width=0.9\columnwidth]{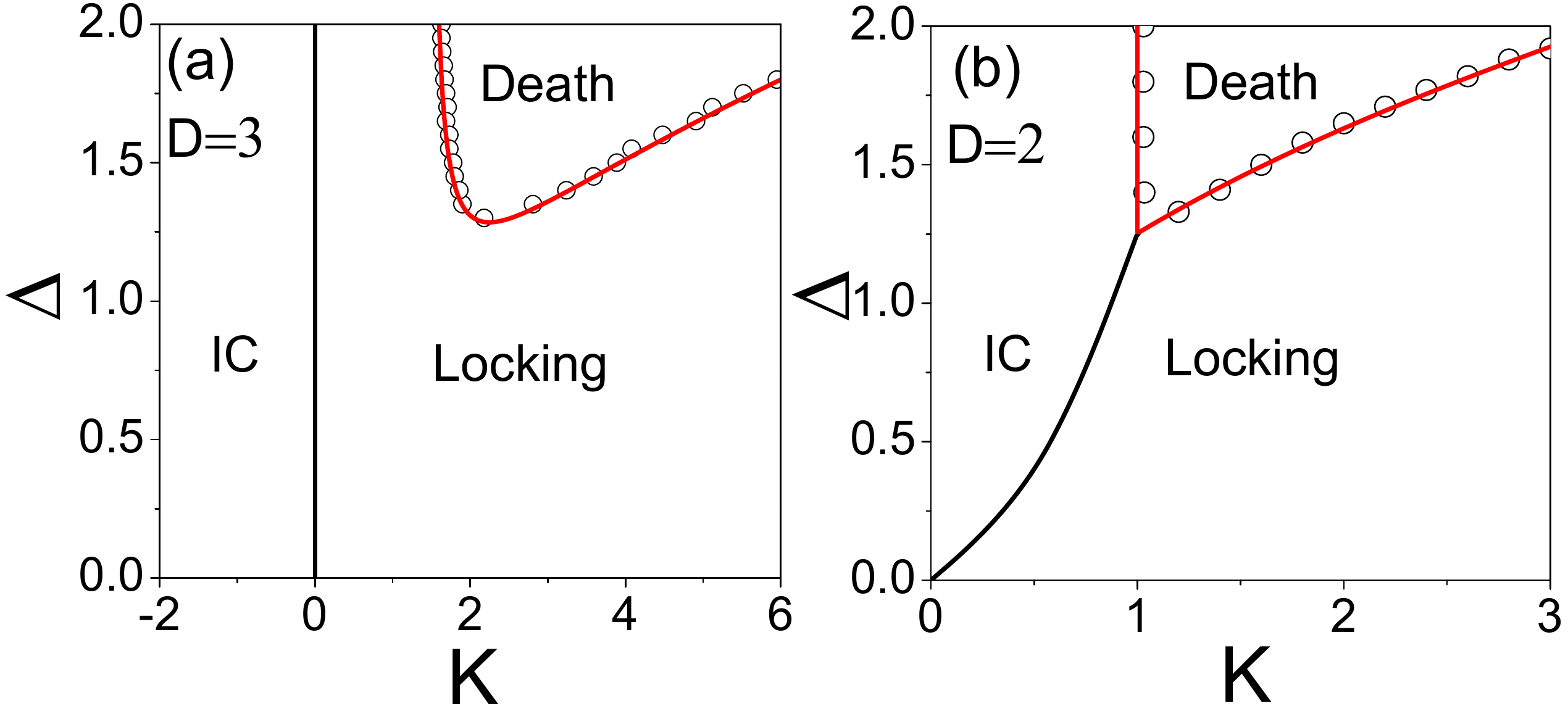}
\end{center}
\centering \caption{Phase diagram for the system (\ref{DSL}) with $D(D-1)/2$ elements of $\textbf{W}_i$ distributed
according to ${Norm}(0,\Delta^2)$ for (a) $D=3$ and (b) $D=2$. The black and red curves denote
the boundaries of incoherence (IC) and amplitude death determined theoretically, whereas the open circles are the simulation
results of the amplitude death boundary by integrating (\ref{DSL}) with $N=5000$.}  \label{fig2}
\end{figure}

For $D=3$, by writing $G(\vec\omega)=g(\omega)U(\hat{\omega})$, the discrete spectrum for amplitude death further reduces to \cite{SM}
\begin{eqnarray}
\label{ADdsb}
h(s-1+K)=\frac{1}{K}
\end{eqnarray}
with the function $h$ defined as in (\ref{dispersionc}). Amplitude death is stable if (\ref{ADdsb}) has only roots with negative real parts for $K>1$.

For example, for $D=3$, if each element of $\vec{\omega}_i$ is picked randomly
according to $\omega_{ji}\sim {Norm}(0,\Delta^2)$ for $j=1,2,3$,
the distribution of the natural rotations can be written as $G(\vec{\omega})=g(w)U(\hat{\omega})$, where
the distribution of the rotation directions is isotropic and the distribution of the magnitudes is
described by $g(\omega)=\sqrt{\frac{2}{\pi}}\frac{\omega^2}{\Delta^3} e^{-\omega^2/(2\Delta^2)}$ \cite{MB}.
By setting  $s=0$ in (\ref{ADdsb}), we obtain the boundary of the stable amplitude death region governed by \cite{SM}
\begin{eqnarray}
\label{ADinterval}
\frac{1}{3(K-1)}+\frac{2(K-1)}{3}\int_0^{\infty} \frac{g(\omega)}{(K-1)^2+\omega^2}d\omega=\frac{1}{K},
\end{eqnarray}
which is represented by the red curve in Fig. \ref{fig2}(a) depicting the phase diagram of the system in the $(K, \Delta)$ parameter space for $D=3$.
Note that the theoretical prediction by (\ref{ADinterval}) is rather accurate and well confirmed by the simulation
results. For comparison, Fig. \ref{fig2}(b) depicts the phase diagram for $D=2$ with $w_i\sim {Norm}(0,\Delta^2)$.
For $D\geq4$, the phase diagrams, qualitatively similar to Figs. \ref{fig2}(a) and \ref{fig2}(b) for odd and even dimensions,
have been corroborated numerically.

\begin{figure}[!h]
\begin{raggedright}
\centering\includegraphics[width=0.9\columnwidth]{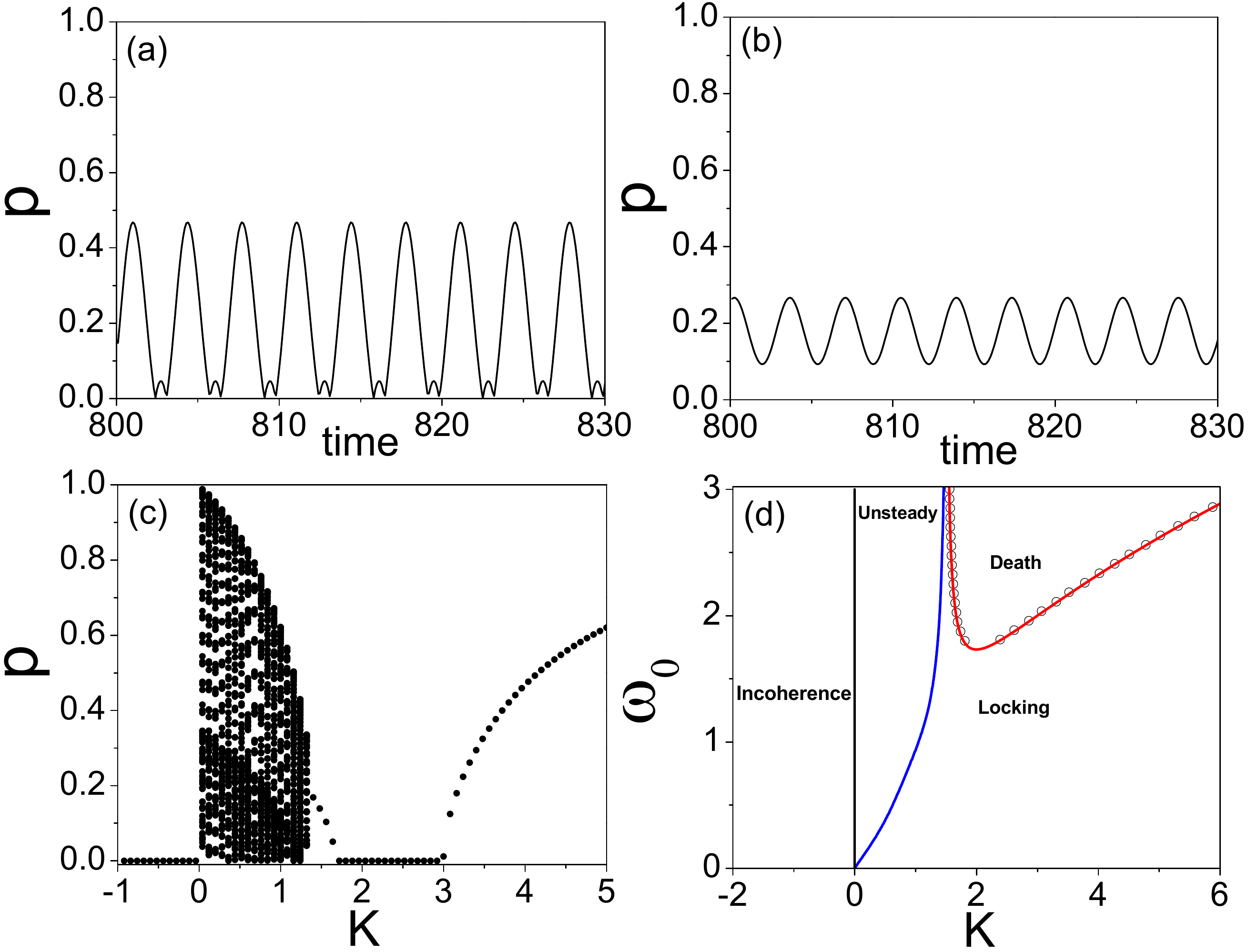}
\par\end{raggedright}
\centering \caption{Results for the system (\ref{DSL}) with $D=3$, where $\vec{\omega}_i$ are now sampled according to
$G(\vec{\omega})=\delta(\omega-\omega_0)U(\hat{\omega})$.
(a) $p(t)$ for $K=1.2$; (b) $p(t)$ for $K=1.36$; (c) $p$ vs $K$. $\omega_0=2$ is fixed. (d) Phase diagram in the parameter space of $(K, \omega_0)$.
The red curve represents the theoretical prediction for the stable coupling interval of amplitude death given by (\ref{ADintervalb}), which is well confirmed by the
numerical results marked by the open circles. The blue curve denotes the unsteady-locking boundary determined numerically.
} \label{fig3}
\end{figure}

For $K>0$, aside from the coherent fixed-point solutions (locked states and amplitude death),  (\ref{DSL}) may also have oscillatory
time-dependent solutions, i.e., the system could display {\it rhythmic states} \cite{Dai2020}, characterized by an unsteady motion
of $p$. For example, Figs. \ref{fig3}(a) and \ref{fig3}(b) show two numerical observations of a time-dependent evolution of $p$ for the system (\ref{DSL}) with $D=3$,
where $\vec{\omega}_i$ are sampled according to $G(\vec{\omega})=\delta(\omega-\omega_0)U(\hat{\omega})$.
Figure \ref{fig3}(c) further depicts $p$ vs $K$. Again, the incoherent state is observed only for $K<0$, and becomes unstable for $K>0$,
which is in accordance with the theoretical prediction of the incoherence.
Interestingly, as $K$ is gradually increased from zero, rhythmic states appear and persist for a large interval of  $K>0$, where
the periodic dynamics of $p$ emerges through the similarly periodic oscillations in the magnitudes of $\vec{r}_i$'s.
After the rhythmic state turning to unstable, the system transits to locked states and amplitude death.
For $G(\vec{\omega})=\delta(\omega-\omega_0)U(\hat{\omega})$ considered above $(\text{i.e.,} \  g(\omega)=\delta(\omega-\omega_0))$, the stable
coupling interval for amplitude death is derived as \cite{SM}
\begin{eqnarray}
\label{ADintervalb}
\frac{\omega_0^2+3-\omega_0\sqrt{\omega_0^2-3}}{3}<K<\frac{\omega_0^2+3+\omega_0\sqrt{\omega_0^2-3}}{3}
\end{eqnarray}
for $\omega_0>\sqrt{3}$. For $D=3$, amplitude death is fundamentally induced by the orientational disorder 
for a large fixed rotation magnitude, which is distinctly different from the case of $D=2$, where amplitude death arises
owe to a sufficiently large spread of the natural frequencies \cite{Matthews1990,Mirollo1990,Matthews1991}.
In fact, amplitude death is impossible to be stabilized for the case of $D=3$ in the absence of orientational disorder \cite{SM}.
For a global view, Fig. \ref{fig3}(d) portrays the phase diagram of the system in the $(K, \omega_0)$ plane.
Clearly, the system (\ref{DSL}) can experience both steady behaviors and
rhythmic states if $\vec{\omega}_i$  sampled according to $G(\vec{\omega})=\delta(\omega-\omega_0)U(\hat{\omega})$ \cite{periodic}.

To conclude, we have introduced and studied a new model of globally coupled $D$-dimensional generalized limit-cycle oscillators with amplitude dynamics.
Under the weak coupling limit $K\rightarrow0$, our model reduces to the $D$-dimensional Kuramoto phase model, which is akin to a similar classic construction of the seminal Kuramoto
phase model from weakly coupled two-dimensional limit-cycle oscillators \cite{Leon2019,Leon2022}.
In this sense, our work puts the recent studies regarding the $D$-dimensional Kuramoto model \cite{Chandra2019,Chandra2019b,Chandra2019c,Dai2020,Kovalenko2021,Dai2021,Lipton2021,Barioni2021,Barioni2021b}
on a stronger footing by providing a much more general framework to consider the previous results,
owing to no longer being constrained by fixed amplitude dynamics.
Thus, our model may find strong potential for actual applications in a wider range of physical, biological, and technological systems involving quenched random
rotation axes and frequencies, such as leading to a deeper understanding of the collective motion in three-dimensional
swarming systems with helical trajectories \cite{Zheng2021},
the spatiotemporal alignment of beating cilia \cite{Niedermayer2008}, the ferromagnetic resonance in biomagnetism \cite{Vonsovskii1966}, etc.
It should be highlighted that the emergence of rhythmic states in the model (\ref{DSL}) strongly depends on the distribution $G(\vec{\omega})$, whose
underlying principles as well as the necessary or sufficient conditions on the distribution $G(\vec{\omega})$ that lead to rhythmic states for $D\geq3$
deserve a detailed study, which would be the scope of future work.
Further, there also lies the possibility of various extensions of our model,
such as including the role of noise, external forces, network-based coupling, etc., which may open up a prosperously new
area of research and will have great impacts in the field of
nonlinear (collective) dynamics and complex systems.


\begin{acknowledgments}
{\it The authors are grateful to the three anonymous reviewers for their
very constructive comments, which helped us to greatly improve this work.}
\end{acknowledgments}



\begin{thebibliography}{99}
\bibitem{perturbations}
The {\it ansatz} for the samll perturbation is of the form $F=F_0+\varepsilon e^{st}\xi(r,\theta,\phi,\vec{\omega})$ $(0<\varepsilon\ll 1)$,
where $s$ is the characteristic exponent used to characterize the growth rate of the perturbation.
If all the real parts of $s$ are negative, the perturbation decays exponentially in time, thus
the incoherent state is stable. In contrast, if $s$ has a positive real part, the incoherent state becomes unstable,
because the perturbation grows exponentially.
\bibitem{ADcsstable}
The continuous spectrum for amplitude death is given by roots of $\text{det}\left((s-1+K)\textbf{I}-\textbf{W}\right)=0$,
where $\textbf{W}$ is an antisymmetric matrix. By applying the fact that the real part of all the eigenvalues of any real antisymmetric matrix is zero,
it can be direclty deduced that the continuous spectrum for amplitude death is stable for $K>1$.
\end{thebibliography}
\end{document}